# The Domain Shift Problem of Medical Image Segmentation and Vendor-Adaptation by Unet-GAN

Wenjun Yan[1], Yuanyuan Wang[1(✉)], Shengjia Gu[2], Lu Huang[3], Fuhua Yan[2], Liming Xia[3], and Qian Tao[4(✉)]

[1] Department of Electrical Engineering, Fudan University, Shanghai, China
yywang@fudan.edu.cn
[2] Department of Radiology, Ruijin Hospital, Shanghai Jiaotong University, Shanghai, China
[3] Department of Radiology, Tongji Hospital, Huazhong University of Science and Technology, Wuhan, China
[4] Department of Radiology, Leiden University Medical Center, Leiden, the Netherlands
Q.Tao@lumc.nl

**Abstract.** Convolutional neural network (CNN), in particular the Unet, is a powerful method for medical image segmentation. To date Unet has demonstrated state-of-art performance in many complex medical image segmentation tasks, especially under the condition when the training and testing data share the same distribution (i.e. come from the same source domain). However, in clinical practice, medical images are acquired from different vendors and centers. The performance of a U-Net trained from a particular source domain, when transferred to a different target domain (e.g. different vendor, acquisition parameter), can drop unexpectedly. Collecting a large amount of annotation from each new domain to retrain the U-Net is expensive, tedious, and practically impossible.

In this work, we proposed a generic framework to address this problem, consisting of (1) an unpaired generative adversarial network (GAN) for vendor-adaptation, and (2) a Unet for object segmentation. In the proposed Unet-GAN architecture, GAN learns from Unet at the feature level that is segmentation-specific. We used cardiac cine MRI as the example, with three major vendors (Philips, Siemens, and GE) as three domains, while the methodology can be extended to medical images segmentation in general. The proposed method showed significant improvement of the segmentation results across vendors. The proposed Unet-GAN provides an annotation-free solution to the cross-vendor medical image segmentation problem, potentially extending a trained deep learning model to multi-center and multi-vendor use in real clinical scenario.

**Keywords:** Domain adaptation, Left ventricle Segmentation, GAN, Unet.

## 1 Introduction

Recent years have witnessed a tremendous boost in computer vision brought by deep neural networks. Deep convolutional neural networks (CNN) have swept almost every image analysis problem since its resurgence [1,2]. For medical image segmentation, the Unet architecture has achieved remarkable success, surpassing human in speed-accuracy balance [3,4].

### 1.1 The Domain Shift Problem of CNN Segmentation

On the other hand, although CNN is argued to be biologically inspired by human vision, researchers have discovered an intriguing gap between human and computer vision. A type of domain variance can be formulated as adversarial perturbation. In tribute to the well-known Panda-Gibbon example [5], we generated adversarial examples for a well-trained Unet for the left ventricle (LV) segmentation task in magnetic resonance image (MRI). We used the fast gradient sign method: the perturbation was computed as equal to the sign of the gradient of the loss function with respect to the input. The perturbation was meant to distract the CNN in a guided manner.

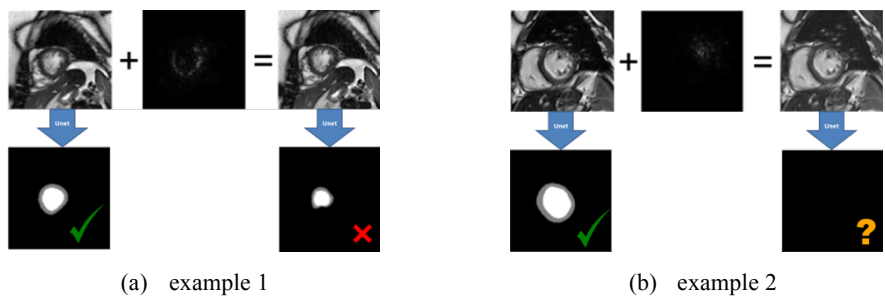

(a) example 1　　　　　　　　　　(b) example 2

**Fig. 1.** Two examples to show that the trained Unet are vulnerable to the carefully calculated perturbation added to the original image. The perturbation hardly affects human vision, but leads to failure of the Unet: in the first example, the segmentation went wrong; in the second example, the segmentation completely failed.

Fig.1 shows two examples of such adversarial attacks. Although almost imperceptible to human eye, the minor perturbation completely misguided a well-trained CNN. In practice, such variance may also come from "distribution shift" [6], which is common when medical images come from different centers, vendors, or acquisition parameters. This leads to a major difficulty of deploying CNN in real clinical scenarios: the performance of CNN can be excellent in the test data from the same distribution, but could drop unexpectedly in unseen data from a different distribution.

Generalization capability of a trained CNN is of paramount importance for its utilization in clinical scenario. An intuitive solution is to include as much data as possible in training, which enlarges the scope of distribution learned by CNN. However, manual annotation of data in large amount, for every new (and unknown) distribution, is practically impossible. Data augmentation in the source domain shows not to be the ultimate solution in closing the generalization gap, since distribution shift is not accidental but systematic [7].

In this work, we proposed a generic framework to address this problem, consisting of (1) an unpaired generative adversarial network (GAN) for vendor-adaptation, and (2) a Unet for segmentation, using the LV segmentation from cine MRI as an example. We integrated the two networks in the design, and name it as *Unet-GAN*.

## 2 Method

**LV-Unet.** A supervised CNN, named LV-Unet, can be trained on the source dataset S that has sufficient annotated LV segmentation. LV-Unet follows an asymmetric encoder-decoder structure as illustrated in Fig.2 (a).

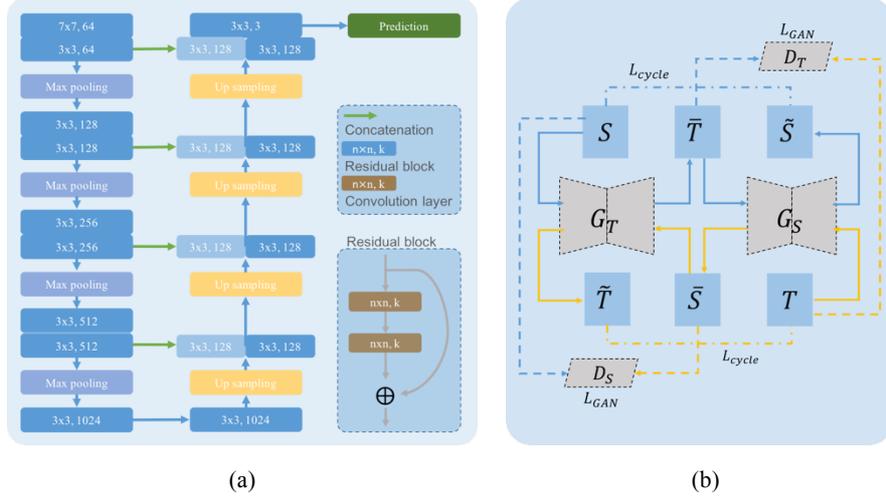

(a)    (b)

**Fig. 2.** (a) The detailed design of the LV-Unet in Unet-GAN, and (b) the basic CycleGAN.

**CycleGAN.** CycleGAN is an established architecture designed for unpaired image-to-image translation [8]. It contains two generators and two discriminators as shown in Fig.2 (b). The generators $G_S$ and $G_T$ play roles of translators between source and target domains. The discriminators $D_S$ and $D_T$ differentiate if the image is original or translated. The loss function is a composite of two losses: one is the general adversarial loss of GAN, defined as: $L_{GAN}(G_x, D_x, x, y) = E_x[\log D(x)] + E_y[1 - \log D(G(y))]$, the other is the $L_2$ loss defined as: $L_{cycle}(G_x, G_y, x, y) = E_{x,y}||x - G_x(G_y(y))||_2$, which enforces the generator to learn the representative features of a specific domain in an unpaired way. The overall loss function of CycleGAN is:

$$L(G_S, G_T, D_S, D_T) = L_{GAN}(G_S, D_S, S, T) + L_{GAN}(G_T, D_T, T, S) + \lambda[L_{cycle}(G_S, G_T, S, T) + L_{cycle}(G_T, G_S, T, S)] \quad (1)$$

**Structural similarity index (SSIM).** The structural similarity (SSIM) index is a measure to assess the quality of generated images [9]. The SSIM formula comprises of three similarity measurements between images *x* and *y*: luminance, contrast, and structure. The three measurements are expressed as follows:

$$l(x,y) = \frac{2\mu_x\mu_y + c_1}{\mu_x^2 + \mu_y^2 + c_1},\ c(x,y) = \frac{2\sigma_x\sigma_y + c_2}{\sigma_x^2 + \sigma_y^2 + c_2},\ s(x,y) = \frac{2\sigma_{xy} + c_3}{\sigma_x\sigma_y + c_3} \quad (2)$$

where $\mu_x/\mu_y$ and $\sigma_x^2/\sigma_y^2$ are mean value and variance of *x* and *y*, respectively, $\sigma_{xy}$ is covariance of *x* and *y*, and $c_i(i = 1,2,3)$ is the parameters to stabilize the dividing operation. The SSIM index is defined as the exponentially weighted combination of the three:

$$SSIM(x,y) = [l(x,y)^\alpha \cdot c(x,y)^\beta \cdot s(x,y)^\gamma] \quad (3)$$

where α, β, γ are the weights of the three terms, respectively.

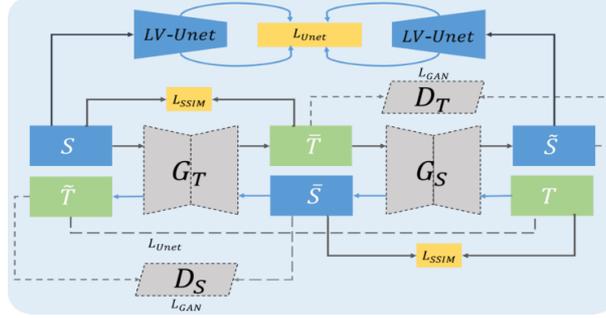

**Fig. 3.** The Unet-GAN architecture, which integrates Unet and CycleGAN.

**Unet-GAN: Dedicated to medical image segmentation across domains.** The proposed Unet-GAN has a loss function combining information at the image and feature level:

*SSIM loss at image level:* The original loss in CycleGAN is an image-level loss computing the mean square error (MSE) between the input image and the translated one. An inherent requirement for medical image domain adaptation is that the luminance and contrast can vary but anatomical structures should be preserved. Based on this observation, we used SSIM as a modified image loss for medical image adaptation: the structure term of SSIM is given a much higher weight than the luminance and contrast terms. $L_{SSIM}$ is defined as: $L_{SSIM} = E_{x,y}||1 - SSIM(x,y)||_2$.

*Unet loss at feature level:* The image-level loss leads to visually similar image, but can still be vulnerable to adversarial noises. To tackle this problem, we propose a novel loss defined at the feature level. The encoder path of the LV-Unet is regarded as an effective feature extractor, highlighting features most relevant for the segmentation task. We used the output from its layers in the encoder path as feature $f$, and measured the MSE of features $f$ between the original and translated image. In this way we enforce the GAN to generate images that produce the same segmentation-specific features. The new loss at feature level is called $L_{Unet}$:

$$L_{Unet} = E_{x,y}||f_x - f_y||_2 = \frac{\sum_{i=1}^{N}(f_{original}^i - f_{translated}^i)^2}{N} \quad (4)$$

The overall loss function of Unet-GAN is defined as:

$$L(G_S, G_T, D_S, D_T) = L_{GAN} + \lambda_1 L_{Unet} + \lambda_2 L_{SSIM} \quad (5)$$

**Overall workflow.** The proposed framework for cross-vendor medical image segmentation consists of three steps:

1. Firstly, the Unet is trained by data from the source domain with sufficient annotation, and the performance in the test set from the same source domain is guaranteed to be up to the state-of-the-art.

2. The generators $G_S, G_T$ and discriminators $D_S, D_T$ in Unet-GAN are trained alternately using unannotated data from both source and target domains, using the loss function defined in the previous section, integrating the Unet features.

3. Finally, data from the target domain are first translated to the source domain by $G_S$, and then fed to the trained Unet for segmentation.

## 3     Experiments & Results

### 3.1     Data

The experiments involved short-axis steady-state free precession (SSFP) cine MR images of 144 subjects acquired by three major MRI machines as three domains (44 Philips samples, 50 GE samples, 50 Siemens samples). Image size varied from 256×256 to 512×512 pixels. All images were rescaled to the same in-plane resolution of 1.5×1.5 mm. Cine MR and label images were cropped at the center to a size of 192×192 for faster training and testing.

Ground truth annotation of the LV myocardium and blood pool were performed on the cine MR images by experienced radiologists. The number of available annotated images in each domain was 4823, 2084, and 2602 for Philips, GE, and Siemens, respectively. Philips data was set as the source domain to train the LV-Unet. We randomly selected 35 subjects out of 44 for training (3920 images) and the rest 9 for testing (903 images). Siemens and GE were defined as the two target domains. To train the Unet-GAN, we separated every domain into training/testing set: 3008/1815 for Philips, 1680/924 for Siemens, and 1320/764 for GE.

### 3.2     Experiments and Performance Evaluation

**LV-Unet:** We used adaptive moment estimation (Adam) optimization with learning rate of 10-4 and a mini-batch size of 10. The number of epochs was set to 35 when network converged and the mean Dice coefficient reached 88% on the test dataset (the state-of-art for LV segmentation [10,11]).

**Unet-GAN:** We used Adam optimization with learning rate of $10^{-5}$. Weighting parameters $\lambda_1$ was set to 15 and $\lambda_2$ was set to 5 empirically. As GAN is generally hard to train, we applied the early stopping strategy during training to prevent model overfitting. Two Unet-GANs were trained to translate image from one domain to another one: 1) Siemens→Philips, trained with 3008 Philips images and 1680 Siemens images, and 2) GE→Philips, trained with 3008 Philips images and 1320 GE images.

Based on our experiments with the LV-Unet, the layers after three max-pooling operations tend to extract high-level semantic features. Therefore, we selected features from the 7th and 9th residual blocks in the LV-Unet for the Unet loss. For the SSIM loss, $\alpha, \beta, \gamma$ were assigned the value of 0.1, 0.1, and 1, respectively.

We performed comparative experiments to evaluate the performance improvement brought by Unet-GAN, with the follow three scenarios tested:

1) Segment data of the target domain directly by LV-Unet trained on the *clean* source domain (i.e. original data);

2) Segment data of the target domain by LV-Unet trained on the *pollute*d source domain (i.e. original data added with random noise);

3) First translate data of the target domain by Unet-GAN to the source domain, then segment the translated data by the LV-Unet trained on the clear source data.

The polluted Unet meant to evaluate how much systematic augmentation can address the domain shift problem, in comparison to the proposed Unet-GAN. The performance of LV segmentation was evaluated in terms of Dice overlap index between the ground truth and the segmentation results.

### 3.3 Results

The Dice indices resulting from the comparative experiment is reported in Table 1. Figure 4 and 5 give some examples, showing the performance degradation caused by changes of domain and the improvement brought by Unet-GAN.

The original LV-Unet trained on the "clean" Philips data appeared sensitive to the change of domain, and the performance drops significantly, especially on Siemens data (from 0.892 to 0.474). When trained on "polluted" Philips data with random noise, the segmentation accuracy degraded on the Philips data itself (from 0.892 to 0.877) but improved on Siemens and GE data with a small margin (0.474 to 0.502, and 0.727 to 0.813, respectively). Fig. 4 and 5 provide a few examples of the vendor-adaptation results for Siemens and GE, respectively. In the first row, it can be seen while LV in the original images can be easily recognized by human eyes, the well-trained LV-Unet from another domain did not perform as anticipated, often having the LV under-segmented. In the second row, the Unet-GAN translated images are shown, and in the third row, the segmentation results on the translated images are shown (overlaid on the original images for easier comparison to the first row). In the Siemens case, brightness changes can be observed in the translated images, while in the GE case, changes are hardly discernable. In both cases, nevertheless, the final segmentation results were substantially improved by the Unet-GAN translation (0.474 to 0.805, and 0.727 to 0.867, respectively).

**Table 1.** Comparison of segmentation performance of LV-Unet on datasets from different domains. Three scenarios as aforementioned are compared: clean Unet, noisy Unet, Unet-GAN.

| Dice | Clean Unet | Noisy Unet | Unet-GAN |
|:---:|:---:|:---:|:---:|
| Philips (test) | $0.892 \pm 0.032$ | $0.877 \pm 0.034$ | — |
| Siemens | $0.474 \pm 0.071$ | $0.502 \pm 0.064$ | $\mathbf{0.805 \pm 0.041}$ |
| GE | $0.727 \pm 0.042$ | $0.813 \pm 0.040$ | $\mathbf{0.867 \pm 0.035}$ |

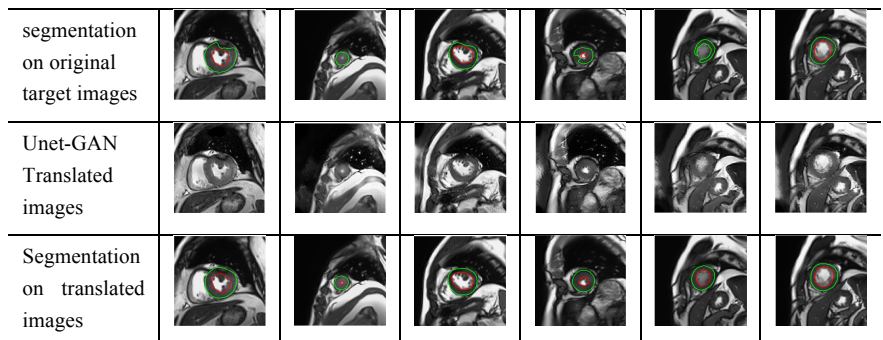

**Fig. 4.** Comparison of LV segmentation results before and after Unet-GAN translation of Siemens examples. The original and translated images were both segmented by the LV-Unet. Upper row: LV segmentation on original target images, middle row: Unet-GAN translated images to the source domain, lower row: LV segmentation on translated images, results overlaid onto the original domain.

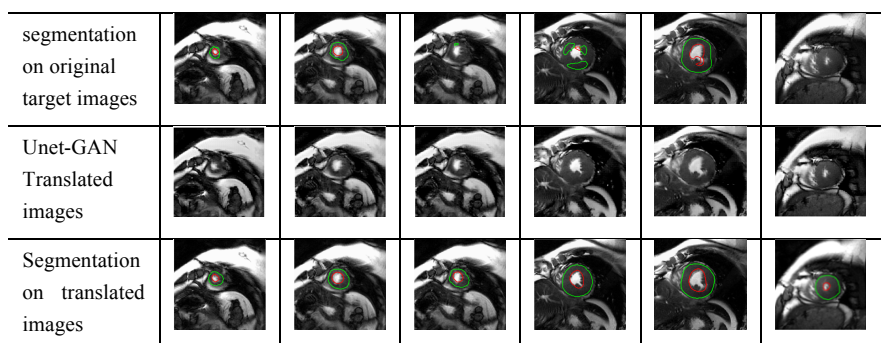

**Fig. 5.** Comparison of LV segmentation results before and after Unet-GAN translation of GE examples. The original and translated images were both segmented by the LV-Unet. Upper row: LV segmentation on original target images, middle row: Unet-GAN translated images to the source domain, lower row: LV segmentation on translated images, results overlaid onto the original domain.

## 4       Discussion and Conclusion

In this paper, we first stated the vulnerability of CNN in medical image segmentation, a practical issue for deployment of CNNs in clinical practice. Intrigued by this problem, we proposed a network architecture combining Unet and GAN, dedicated to the medical image segmentation problems across domains (MRI vendors in this case).

The domain shift problem of CNN has been an inspiring research area in machine learning, and most studies have focused on classification, (e.g., with random-appearing noise, the labelling of an image can unexpectedly change). The domain shift problem of segmentation, however, has not caught specific attention. Although Unet has achieved tremendous success in medical image segmentation, we observed similar phenomenon when a well-trained Unet is applied to data from another domain: even though the images in a target domain are visually good-quality as those in the source domain, the Unet can fail in an unpredicted way. This can be at least partially explained by the independent and identically distribution (i.i.d.) assumption of statistical learning: the CNN can learn the statistical distribution from the available source domain very well, however, if the distribution differs from it in the target domain, it cannot generalize. By the proposed Unet-GAN, the distribution in the target domain can be shifted to fit the original distribution of the source domain, in such a way that it particularly maintains the segmentation performance because the GAN enforces similarity of segmentation-specific Unet features.

In conclusion, we proposed a network architecture called Unet-GAN, which adapts images across vendors such that a Unet trained on one vendor can generalize well to another vendor without need of new annotation. The method extends the utilization of a trained CNN to multi-center and multi-vendor use in real clinical scenario.